# Investigation Solvation Dynamics and Isomerization of Dye IR-140 Using pump supercontinuum-probing Technique

Danling Wang, Hongbing Jiang, Songjiang Wu, Hong Yang, Qihuang Gong *

*State Key Laboratory for Mesoscopic Physics & Department of Physics, Peking University, Beijing 100871, China*

Junfeng Xiang[1], Guangzhi Xu[2]

1) *Technical Institute of Physics and Chemistry, the Chinese Academy of Sciences, Beijing 100101, China*

2) *State Key laboratory for Structural Chemistry of Stable and Unstable Species, Institute of Chemistry, The Chinese Academy of Sciences, Beijing 100080, China*

Abstract:

The solvation dynamics and isomerization process of an organic dye, IR-140 (5,5'-dichloro-11-diphenylamine-3,3'-diethyl-10,12-ethylene-thiatricarbocyanine-perchlorate), in polar solvents and nonpolar solvents have been investigated using pump supercontinuum-probing (PSCP) technique. In all solvents, the dynamics exhibits solvent-dependent. Solvent induced spectral shifts of the absorption and emission spectra of IR-140 have also been studied in a variety of solvents. At the same time, the photoisomerization process has been examined. The approximate energy-band structure of IR-140 was also supposed firstly.

# I. Introduction

Photophysical properties of triacarbocynine dye in solvents have motivated a large amount of scientific work recent years[1~5]. 5,5'-dichloro-11-diphenylamine-3,3'-diethyl-10,12-ethylene-thiatricarbocyanine-perchlorate(IR-140) as one kind of triacarbocynine dye, the photophysical properties and photoisomerization processes have also considerably gained interesting[6~8]. Especially, the photoisomerization of IR-140 in ethanol and in DMSO had been firstly studied by J.P.Fouassier et.al in 1977[6]. In 1998, Y.H.Meyer et.al was explained photoisomer process and forecasted the excited-state absorption of IR-140 by global spectral analysis [1]. Then recently, J.C.Gumy et.al studied the solvation dynamics of IR-140 in polar solvents using the femtosecond transient grating technique[8]. They observed the solvation dynamics was dependent on the probe wavelengths: at 830nm, the diffusive motion is major element while at 780nm the dynamics is dominated by the inertial motion (intramolecular charge transfer process).

In this paper we report the pump supercontinuum-probing (PSCP) measurements on the transient absorption spectrum and transient relaxation of IR-140 in polar solvents and nonpolar solvents with a time resolution of 130fs and probe wavelength ranging from 400nm to 1000nm. As the femtosecond PSCP technique has broad spectral coverage and high time resolution [9-11], more information about the energy structure of IR-140 can be obtained from its transient spectra. Meanwhile, the photophysical properties about the probe wavelength dependence of the dye IR-140 in variational solvents is fully proved. More information about the photoismerization process of IR-140 was obtained from its transient spectra and more detailed energy structure of IR-140 was firstly reported.

## II. Experimental Setup

The experimental setup for PSCP was based on a Ti: Sapphire chirped-pulse amplification *CPA* system. The laser system (TSA-10, Spectra-Physics, USA) delivered 807nm wavelength laser pulses with repetition rate of 10 Hz and pulse duration of 130 fs. The maximum pulse energy was up to 10 mJ. The experimental setup for measuring the optical transient relaxation of the dye IR-140 is shown in Fig.1. The laser pulse was split into two parts by a beamsplitter (*BS*). The transmitted beam, after passing a variable optical delay line (ODL) controlled by a computer, was used for optical pumping. The pump beam had a diameter of 2mm on the sample and the pulse intensity can be varied from 0.1mJ to 10uJ by rotating a half-wavelength plate in front of a polarizer. The reflected beam, with an energy of about 1.0mJ, was focused with a 6.5-cm lens (*L1*) into distilled water to generate supercontinuum probe pulses. The center part of the supercontinuum, with an energy of 1.0uJ, was converged by dispersion-free concave mirror with metal coating and reflected by a fused silica plate (*P*) with thickness of about 4-mm. The reflected probe beam from the front surface, called as probe beam 1, was focused to intersect with the pump beam in the sample solution and to detect dynamic absorption changes caused by the pump beam. The reflected probe beam from the rear surface of the plate, probe beam

2, was used as the reference to monitor the energy fluctuation of probe 1. The sample solution with concentration of $1\times10^{-5}$ M was filled in a 2mm thickness glass cell resulting in the initial absorption about 50% at 807nm. After interaction with the sample, the transmitted probe beam were input into a fiber spectrometer (PC2000, Ocean Optics, Inc.) and displayed in the computer. The transient absorption spectroscopy was detected by changing the optical delay of the probe with respected to the pump.

## III. Results and discussions

The molecular structure of IR-140 is shown in Fig. 2. It is a triacarbocynine dye with extended polymethine chains and exhibit absorption and emission maxima in the near-infrared(NIR).

The steady-state absorption spectra were measured in polar solvents such as alcoholic solvents and nonpolar solvents such as chloroform, tetrahydrofuran etc. which is shown in Table 1. Noticeably, the position of the absorption maximum is shifted to shorter wavelengths with increasing the polarization of the solvents. That is, the maximal absorption is 799 in methanol while shifts to 820nm in Benzene. On the other hand, the emission spectra of IR-140 in ethanol which excited at 810nm can be deconvoluted into two components: one centered at 823nm and the other centered at 840nm which shown in Fig.3. The similar results can be obtained in other solvents.

**Table I  The steady state absorption of IR-140 in varied solvents**

| Parameter  Solvents | Refraction index (n) | Dielectric constant($\varepsilon$) | $\lambda_A$(maxima, nm) |
|---|---|---|---|
| Methanol | 1.3286 | 32.35 | 799 |
| Ethanol | 1.3616 | 25 | 803 |
| Propan-1 | 1.3856 | 20.81 | 806 |
| n-Butyl alcohol | 1.3992 | 17.4 | 807 |
| Tetrahydrofuran | 1.4050 | 7.52 | 812 |
| Chlorobenzene | 1.5597 | 5.45 | 823 |
| Benzene | 1.5011 | 2.283 | 820 |

Thus, the spectra shifts and the two components in the emission spectra can be attributed to existence of conformation change according to the Ref.[12] and Ref.[3]: one is the all-*trans* and the other is mono-*cis*.  In order to analyze the photophysical properties of IR-140 in the round, we investigated the transient relaxation and transient spectra of IR-140 in different solvents by PSCP technique. When near infrared pump pulses 807nm resonantly excited the sample, the time-resolved signals as a function of the delay time and probe wavelength can be investigated. In our pervious paper[13], from the transient absorption and gain spectra of IR-140 in ethanol at different time intervals, we reported the negative signal ( $650nm > \lambda > 500nm$ ) - the first excited-state absorption (ESA) was corresponding to the transition from the $S_1$ to $S_n$. While according to the positive signal(for $900nm > \lambda > 720nm$ ), we have hypothesized there existed an isomerization process in the wavelengths covering from 720nm to 850nm. Now we compared the transient transmission spectra of IR-140 in polar solvent (such as ethanol) and in nonpolar solvent (such as benzene) at different time intervals in Fig.4. The positive signal exhibit red shift and there is a new absorption band appearing which centers around at 820nm in benzene while it is not obvious in ethanol(denoted by the arrow in Fig.4). Considering the new absorption band in the nonpolar solvent-benzene, we can mainly ascribe the spectroscopic difference to the different isomerization rates between the *cis* and *trans* conformation of IR-140 in thus two kinds of solvents when we ignore the intersyestem crossing in triacabocyanine dye and triplet-triplet absorption of the first singlet excited state[2,14] because of the low quantum efficiency($<10^{-3}$). The investigation shows the different of isomerization rates may mainly originate from following effects. First, because the isomeriazation process involves torsional motion about the polymethine chain, according to J.P.Fouassier et.al[4], but the polymethine chain is easier to rigid with increasing the polarity of the solvents then result in the nonradiative efficiency of IR-140 will be lower than in the low-polar solvents[2]. That is, the isomerization rates will be increased with decreasing the polarization solvents. The second effect might be due to ion-pairing between the dye and its counterion[3,15]. In low-polarity solvents, the dye, bearing an overall negative charge, will most likely be paired with positively charged counterion. This should force the equilibrium toward the all-*trans* conformation while in polar solvents the unpaired dye would adopt mono-*cis* conformation. In a word, all of the effects show that the isomerization rate of IR-140 is lower in the more polar solvents which is consistent with the results that isomerization efficiency of another tricabocyanine dye DCM is very low in the more polar solvent[16]. According to the authors such as West et.al[17], the blue-absorbing species could be assigned to the *cis* isomer and red-absorbing one to the all-*trans*.

Thus, we can tell from our transient spectra in Fig.4. that the new absorption band in benzene should be assigned as the *cis* isomer. Therefore, we can explain the higher rate of the *cis* isomer in the nonpolar solvent resulting in the spectroscopic difference of IR-140 which wavelengths ranging from 720nm to 850nm.

Then considering the red-shift of the absorption spectra in nonpolar solvents compared to polar solvents, according to D.Noukakis et.al[3] and J.C.Gumy et al[8], we can conclude that the position of the absorption maximum is shifted to longer wavelengths by dispersive interactions between the transition dipole of the dye and the electronic polarizability of the solvent. And the different proportion between the inertial motion (intramolecular charge transfer process) and diffusional motion can also result in the spectral shifts. Because the solvent configuration for the red side involve more diffusional motion than the configurations corresponding to blue side absorption. Thus can explain why the absorption band of IR-140 is shifted to a shorter wavelength as the polarity of the solvent increases.

Linking with our pervious paper[13] and comparing the transient dynamics of IR-140 in polar solvents with in nonpolar solvents. The relaxation can also be fitted by double-exponent process and the delay time is probe wavelength-dependent no matter what the solvent is polar or nonpolar. At the same time, the generating mechanism of ESA (excited-state absorption) and SE (stimulated emission) is dominated by linear absorption of 807nm which was already proved in our pervious paper[13]. But the delay time is observed to be also solvent-polarity dependence except for the wavelength-dependent in this paper. For example, in benenze, the time constant at 830nm is $\tau_1 = 0.99 \pm 0.14(ps), \tau_2 = 0.13 \pm 0.07(ns)$; In tetrahyadrofuran, the delay time is shorter than benenze because of its dielectric constant smaller than benenze if we ignored the effect of viscosity (the viscosity of benenze is 0.604 while the tehrahyadrofuran is 0.456). On average, the delay time in low-polar solvent is shorter than in the alcoholic solvents, which is content with the report that Jean-Pierre Fouassier et al[6] observed the fluorescence lifetime was shortened in ethanol ($\varepsilon = 25.00$) than in DMSO($\varepsilon = 46.7$). The contrast of the time constant at 840nm in different kinds of solvents is shown in Table 2.

**Table 2 The delay time at 840nm in different solvents**

| Parameter<br>Solvents | Fitting results $\tau_1$ (ps) | Fitting results $\tau_2$ (ns) |
|---|---|---|
| Methanol | 2.14 ± 0.09(A$_1$=0.76 ± 0.20) | 0.58 ± 0.07(A$_2$=0.24 ± 0.01) |
| Ethanol | 1.72 ± 0.12(A$_1$=0.38 ± 0.02) | 0.27 ± 0.02(A$_2$=0.62 ± 0.01) |
| 1-Butyl alcohol | 1.52 ± 0.03(A$_1$=0.17 ± 0.20) | 0.17 ± 0.06(A$_2$=0.83 ± 0.03) |
| Tetrahydrofuran | 1.40 ± 0.08(A$_1$=0.15 ± 0.20) | 0.15 ± 0.03(A$_2$=0.85 ± 0.01) |
| Benzene | 0.99 ± 0.14(A$_1$=0.06 ± 0.02) | 0.13 ± 0.07(A$_2$=0.94 ± 0.01) |

The different delay time in varied solvents can also prove the isomerization process of IR-140. That is, the fluorescence lifetime is shortened with increasing the rate of the isomerization.

In a word, we obtain the energy-band structure of IR-140 in more detail which is shown in Fig.5. according to the solvation dynamics and isomerization process of IR-140. In Fig.5, when the 807nm laser pulses excited linearly the IR-140, the excited-state absorption was generated from $S_1$ to $S_n$ centering at 550nm. At the same time, the first excited singlet state is twisted by rotating the polymethine chain which caused trans→cis isomerization process and the center of isomer absorption was about at 820nm. According to the spectral shifts and the different dynamics in different solvents, we can tell that the cis isomer conformation is dominant by the wavelengths in the range from 820nm to 900m while the wavelengths ranging from 720nm to 820nm is corresponding to the molecular conformation of the all-trans.

## IV. Conclusion

In summary, the solvation dynamics and isomerization process of the dye IR-140 were investigated by using pump supercontinuum-probing(PSCP) technique with 130fs time resolution and probe wavelengths from 400nm to 1000nm. We observed the different transient spectra and spectral shift in polar and nonpolar solvents. The solvent-dependent transient spectra is considered to be due to tran-cis isomerization in the first excited singlet state twist form because of the rotation of the polymethine chain and ion-pairing effect. On the other hand, the spectral shift can be ascribed to dispersive interactions between the transition dipole of the dye and the electronic polarizability of the solvent and different solvent configurations with different dynamics. The transient relaxation between the two kinds of solvents was also investigated at the same time. In this paper, we not only approved the results in our pervious paper but also obtained the more detailed energy band of the IR-140 based on the spectral shifts and photoisomerization process. All of above investigation provide us an ingelligible image about the photophysical process of IR-140 and also help us understand clearly the photophysics properties, photoisomerization and dye-solvent actions in the other large organic tricabocyanine dyes.

*Acknowledgements*

The author is grateful to Prof. Zongju Xia for assistant in the experiment. The work was supported by the National Key Basic Research Special Foundation (NKBRSF) under Grant No.G1999075207 and National Natural Science Foundation of China under grant No. 19884001 and 19525412.

## *Picture Captions*

Fig.1 The experimental setup of Pump supercontinuum-probe

Fig.2 The molecular formulae of IR-140

Fig.3. The Steady state absorption and emisson spectra of IR-140

Fig.4(a) Transient spectra at –0.7ps and 8.0ps from 400nm to 1000nm in ethanol

Fig.4(b) Transient spectra at –2.0ps and 8.0ps from 400nm to 1000nm in benzene

Fig.5.The energy band of IR-140

Fig.1 D.L.Wang el.al

Fig.2 D.L.Wang et.al

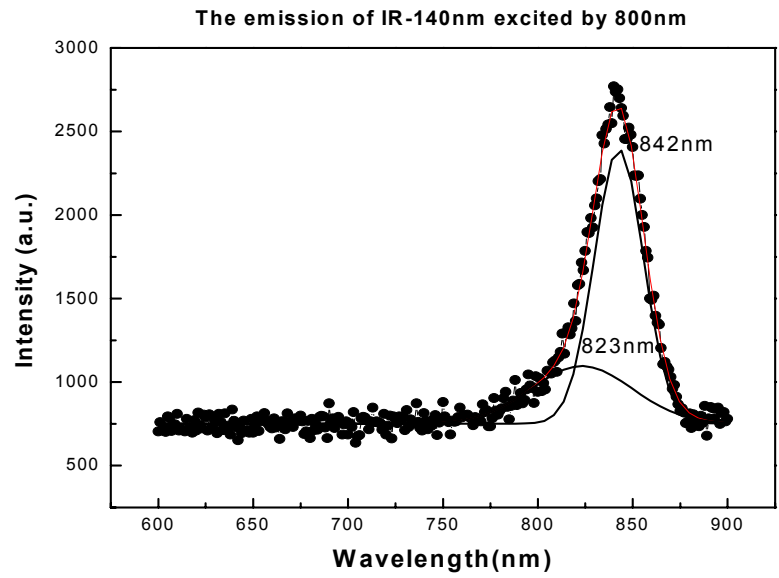

Fig.3    Wang D.L et.al

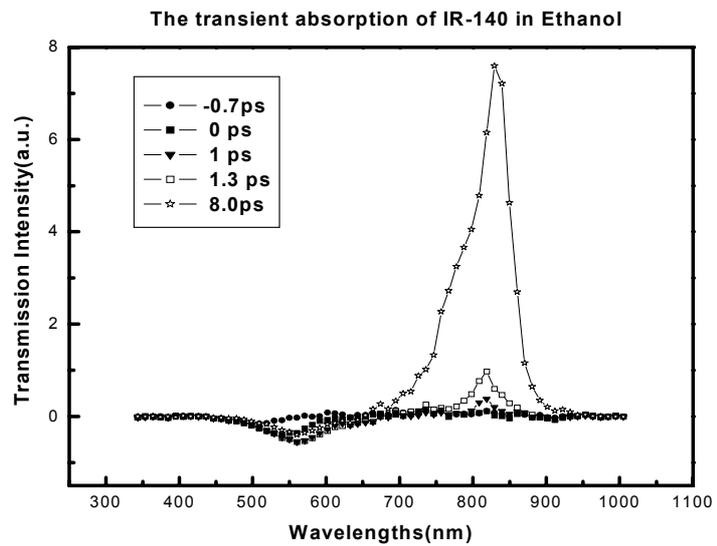

Fig. 4(a)    D.L.Wang et.al

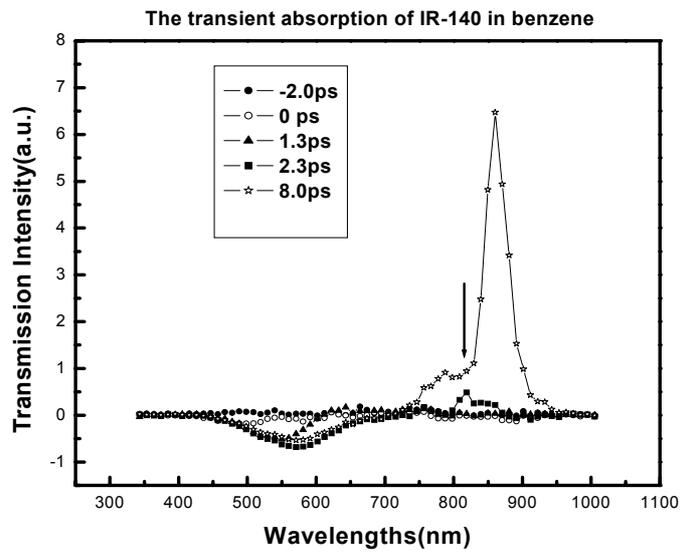

**Fig.4(b) Wang D.L et.al**

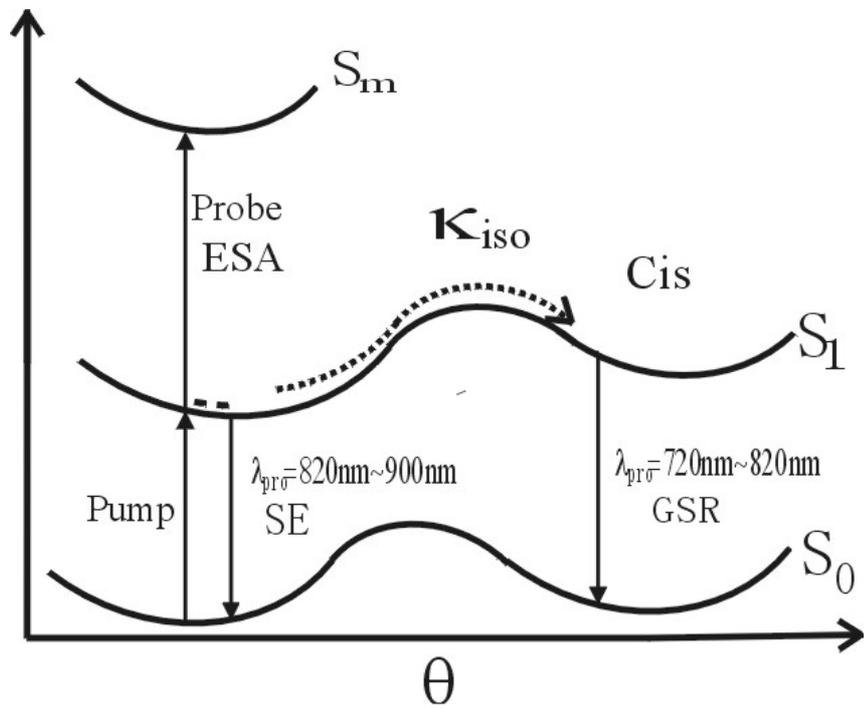

**Fig.5 Wang D.L et.al**